\documentclass[12pt,showpacs,showkeys,amsmath,amssymb]{revtex4}
\usepackage{amsmath,amsfonts,amsthm,amscd,amssymb,latexsym}
\usepackage{bm}
\usepackage{dcolumn}
\usepackage{graphicx}
\usepackage{epstopdf}
\usepackage{color}
\usepackage{epsf}
\usepackage{epsfig}
\usepackage{graphicx, epic, eepic, color}

\def\a{\alpha}
\def\b{\beta}

\def\d{\delta}

\def\@{\partial_}

\def\negenspace{\kern-1.1em}

\def\sqr#1#2{{\vcenter{\hrule height.#2pt\hbox{\vrule width.#2pt
height#1pt \kern#1pt \vrule width.#2pt}\hrule height.#2pt}}}
\def\square{\mathchoice\sqr64\sqr64\sqr{4.2}3\sqr{3.0}3}

\begin{document}
\title{Linearized Gravitational Waves in Nonlocal General Relativity}

\author{C. Chicone}
\email{chiconec@missouri.edu}
\affiliation{Department of Mathematics and Department of Physics and Astronomy, University of Missouri, Columbia,
Missouri 65211, USA }

\author{B. Mashhoon}
\email{mashhoonb@missouri.edu}
\affiliation{Department of Physics and Astronomy,
University of Missouri, Columbia, Missouri 65211, USA}

\begin{abstract} 
We investigate gravitational radiation in the linear approximation within the framework of the recent nonlocal generalization of Einstein's theory of gravitation. In this theory, nonlocality can simulate dark matter; in fact, in the Newtonian regime, we recover the phenomenological Tohline-Kuhn approach to modified gravity. To account for the observational data regarding the rotation curves of spiral galaxies, nonlocality is associated with a characteristic length scale of order $\lambda_0$~=~10 kpc. It follows that in nonlocal gravity, the treatment of extremely low-frequency ($\sim 10^{-12}$ Hz) gravitational waves with wavelengths of order $\lambda_0$ would be quite different than in general relativity.  However, for radiation of frequency $\gtrsim 10^{-8}$ Hz, which is the frequency range that is the focus of current observational searches, the corresponding wavelengths are very small compared to $\lambda_0$. We find that in this frequency regime the nonlocal deviations from general relativity essentially average out and can be safely neglected in practice. 
\end{abstract}



\maketitle

\section{Introduction}

In special relativity, Lorentz invariance is extended to actual (accelerated) observers in a pointwise manner~\cite{1}. This \emph{locality} assumption originates from Newtonian mechanics~\cite{2, 3, 4, 5a, 5b}. However, as pointed out by Bohr and Rosenfeld~\cite{6, 7}, field measurements cannot be performed instantaneously; indeed, such determinations involve an average over the past history of the accelerated observer. The conflict is resolved in \emph{nonlocal special relativity}, where the acceleration of the observer's world line explicitly appears in a universal kernel that acts as the weight function for the observer's memory of its past acceleration~\cite{8}. 

The principle of equivalence implies a deep connection between inertia and gravitation. It is therefore natural to develop a nonlocal generalization of Einstein's general relativity (GR). This has been achieved via the teleparallel equivalent of general relativity (GR$_{||}$), which is a gauge theory of the Abelian group of spacetime translations---see~\cite{9, 10, 11, 12, 13, 14} and the references cited therein. In fact, there is some formal resemblance between GR$_{||}$ and  electrodynamics. In electrodynamics, the constitutive relations between $(\mathbf{E}, \mathbf{B})$ and $(\mathbf{D}, \mathbf{H})$ could be nonlocal~\cite{15, 16}; in a similar way, one can introduce nonlocality into GR$_{||}$ via constitutive kernels. The simplest possibility would involve a \emph{scalar} constitutive kernel that acts as a weight function for a certain average over past events. These are connected to the present through the \emph{world function}~\cite{Sy} in nonlocal general relativity~\cite{17, 18, 19, 20, 21}. 

Nonlocal general relativity, or equivalently nonlocal GR$_{||}$, is a tetrad theory of gravity, where the gravitational potentials are defined via the tetrad field $e_{\mu}{}^{\hat{\alpha}}(x)$ and $x$ here represents an event in spacetime with coordinates $x^\mu=(ct_x, \mathbf{x})$. The tetrad frame is orthonormal, namely, 
\begin{equation}\label{1}
 e_\mu{}^{\hat{\alpha}} e_\nu{}^{\hat{\beta}} \eta_{{\hat{\alpha}} {\hat{\beta}}}=g_{\mu \nu}\,,
\end{equation}
where $g_{\mu \nu}$ is the metric tensor of the Weitzenb\"ock spacetime and the Minkowski metric tensor  $\eta_{{\hat{\alpha}} {\hat{\beta}}}$ is given by diag$(-1,1,1,1)$ in our convention. In this paper, Greek indices run from 0 to 3, while Latin indices run from 1 to 3. The hatted Greek indices ${\hat{\alpha}}$, ${\hat{\beta}}$, etc., refer to \emph{anholonomic} tetrad indices, while $\mu$, $\nu$, etc., refer to \emph{holonomic} spacetime indices. We use units such that $c=1$, unless otherwise specified; moreover, we set $\kappa=8 \pi G/c^4$. 

The geometric framework of nonlocal gravity is based on Weitzenb\"ock spacetime in which the tetrad frame field is globally teleparallel; that is, the spacetime is a parallelizable manifold. We can therefore choose a  global Cartesian tetrad frame for which, in addition to the curvature of the Weitzenb\"ock spacetime, the corresponding connection vanishes as well. \emph{We will work with such a tetrad frame and the corresponding gravitational potentials throughout this paper.} In general, the tetrad field represents the gravitational and inertial potentials in the Weitzenb\"ock spacetime of nonlocal gravity. Our particular choice of tetrad frame in effect eliminates the inertial effects and renders the resulting global Cartesian tetrad field purely gravitational in nature~\cite{14}. 

The \emph{gravitational field strength} in nonlocal gravity is given by 
\begin{equation}\label{2}
 C_{\mu \nu}{}^{\hat{\alpha}}=\partial_{\mu}e_{\nu}{}^{\hat{\alpha}}-\partial_{\nu}e_{\mu}{}^{\hat{\alpha}}\,,
\end{equation}
which is the torsion of the Weitzenb\"ock spacetime in our convention. It is interesting to compare this equation with the standard expression for the electromagnetic field strength in terms of the gauge potential. A particular combination of the field strength and the tetrad frame field, designated as the \emph{modified field strength}, plays an important part in GR$_{||}$ as well as its nonlocal generalization and is given by
\begin{eqnarray}\label{3}
\mathfrak{C}_{\mu \nu}{}^{\hat{\alpha}} :=\frac 12\,
    C_{\mu \nu}{}^{\hat{\alpha}} -C^{\hat{\a}}{}_{[\mu \nu]}+2e_{[\mu}{}^{\hat{\a}} C_{\nu]{\hat{\b}}}{}^{\hat{\b}}\,.
\end{eqnarray}
The field equations of nonlocal general relativity may then be expressed as
\begin{equation}\label{4}
 \partial_{[\mu} C_{\nu \rho]}{}^{\hat{\alpha}}=0\,,
 \end{equation}
\begin{equation}\label{5}
  \partial_\nu{\cal H}^{\mu \nu}{}_{\hat{\alpha}} =\sqrt{-g}~(T_{\hat{\alpha}}{}^\mu + E_{\hat{\alpha}}{}^\mu)\,.
\end{equation}
With our conventional choice of Cartesian tetrad frame, Eq.~\eqref{4}, which is the exact analog of the source-free part of Maxwell's equations, is automatically satisfied. We are thus left with our main   field equation~\eqref{5}, in which the excitation ${\cal H}^{\mu \nu}{}_{\hat{\alpha}}$ is related to the modified field strength $\mathfrak{C}^{\mu \nu}{}_{\hat{\alpha}}$ via a ``constitutive" relation. In GR$_{||}$, this constitutive relation is a linear proportionality; that is, ${\cal H}^{\mu \nu}{}_{\hat{\alpha}}= (\sqrt{-g}/{\kappa})\mathfrak{C}^{\mu \nu}{}_{\hat{\alpha}}$. In our nonlocal ansatz, however, this simple constitutive relation is extended to include a nonlocal part that involves a causal scalar kernel ${\cal K}$, as described in detail in Refs.~\cite{17, 18, 19, 20, 21}. The field strength $C_{\mu \nu}{}^{\hat{\alpha}}$, the modified field strength $\mathfrak{C}_{\mu \nu}{}^{\hat{\alpha}}$ and the excitation ${\cal H}_{\mu \nu}{}^{\hat{\alpha}}$ are all antisymmetric in their first two indices. The nonlocal gravity theory clearly specifies the various quantities upon which the kernel could depend~\cite{18}; in addition, we expect that the nonlocal kernel---as the function weighing past events for the present---would vanish for events that are sufficiently distant in time and space. Moreover, $T_{\hat{\alpha}}{}^\mu$ is the matter energy-momentum tensor, while  $E_{\hat{\alpha}}{}^\mu$ is the energy-momentum tensor of the gravitational field, defined by
\begin{equation}\label{6}
\sqrt{-g}~ E_{\hat{\alpha}}{}^\mu:=-\frac 14  e^\mu{}_{\hat{\alpha}}(C_{ \nu \rho}{}^{\hat{\beta}}
{\cal H}^{\nu \rho}{}_{\hat{\beta}}) + C_{{\hat{\alpha}} \nu}{}^{\hat{\beta}} {\cal H}^{\mu \nu}{}_{\hat{\beta}}\,.
\end{equation}
The connection between this \emph{tensor} in GR$_{||}$ and the energy-momentum \emph{pseudotensor} of the gravitational field in GR has been clarified in Ref.~\cite{AGP}. The main issue here is the separation of gravitational and inertial effects~\cite{14}. In GR, the gravitational and inertial contributions are inextricably mixed together in the energy-momentum pseudotensor. In GR$_{||}$, on the other hand, it is possible to define a purely local gravitational energy-momentum tensor. This tensor plus a contribution from the coupling of gravitational and inertial effects then constitutes the pseudotensor. In the present paper, however, inertial effects are absent in our global Cartesian tetrad field. A more complete treatment of the gravitational energy-momentum tensor is contained in Chap. 10 of Ref.~\cite{14}. 

Some of the physical consequences of nonlocal gravity have been explored in Refs.~\cite{17, 18, 19, 20, 21}. Nonlocality can act like \emph{dark matter}; in fact, nonlocal gravity involves a galactic length scale $\lambda_0=10$ kpc in order to provide a satisfactory explanation for the flat rotation curves of spiral galaxies. At present, the observational implications of nonlocality for gravitational physics in the solar system are essentially negligible, since the size of the solar system is very small compared to $\lambda_0$~\cite{19}.  

     The main field equation of nonlocal gravity in its linearized form is presented in the following section. As in general relativity, the general linear approximation to nonlocal gravity can be used to study nonlocal Newtonian gravity as well as gravitational radiation. The nonlocal modification of Poisson's equation of Newtonian gravitation theory has been recently studied in detail in Ref.~\cite{21}. Therefore, the present paper is devoted to the treatment of linearized gravitational waves in nonlocal general relativity. As the treatment of linearized gravitational waves in GR is well known, in this paper we concentrate instead on an examination of the nonlocal deviations of the theory from the standard general relativistic analysis. 

Gravitational radiation damping can explain the steady orbital decay rate of the Hulse-Taylor binary pulsar as well as similar relativistic binary systems; indeed, this circumstance provides \emph{indirect} evidence for the existence of gravitational waves within the context of GR \emph{and} GR$_{||}$~\cite{SS, SSW, MGH}. Moreover, nonlocal gravity involves a galactic length scale that is very much larger than the orbital size of a relativistic binary pulsar, so that nonlocal effects are likely to be negligibly small. Therefore, we expect that the compatibility of the gravitational radiation damping with GR$_{||}$~\cite{SS, SSW, MGH} would still be true in the nonlocal generalization of GR$_{||}$ in much the same way as the influence of nonlocality on gravitational physics in the solar system is completely negligible at present~\cite{19}.  Direct searches for gravitational waves continue at present---for a recent review see Ref.~\cite{Ri}.    

The plan of this paper is as follows. In section II, we present the general linear approximation of nonlocal GR$_{||}$. In this case, the Newtonian limit was investigated in detail in Ref.~\cite{21} and the results are briefly described in section III. The treatment of linearized gravitational radiation within the framework of nonlocal gravity depends upon the use of appropriate nonlocal kernels. These are derived in section IV and applied to the discussion of generation and propagation of gravitational waves in sections V--VII. The length scale $\lambda_0$ associated with nonlocal gravity corresponds to gravitational waves of frequency $\sim 10^{-12}$ Hz, while present observational possibilities involve frequencies $\gtrsim 10^{-8}$ Hz. Under these conditions, we show that the consequences of the nonlocal theory for linearized gravitational radiation turn out to be essentially the same as in GR.     

\section{Nonlocal Gravity: General Linear Approximation}

In the linear approximation, nonlocal GR$_{||}$ can be treated on a background global Minkowski spacetime; that is, the corresponding Weitzenb\"ock spacetime may be regarded as a slightly perturbed Minkowski spacetime. The deviation of the Cartesian tetrad field from the global background inertial axes may therefore be written as
\begin{equation}\label{7}
 e_\mu{}^{\hat{\alpha}}={\d}_\mu ^{\hat{\alpha}}+\psi^{\hat{\alpha}}{}_\mu\,, \quad  e^\mu{}_{\hat{\alpha}}=\d^\mu _{\hat{\alpha}} -\psi^\mu{}_{\hat{\alpha}}\,.
\end{equation}
We assume that the absolute magnitudes of the nonzero components of  $\psi_{\mu {\hat{\alpha}}}$ are so small in comparison to unity that the linear weak-field approximation is valid, in which case the distinction between holonomic and anholonomic indices disappears as well. Henceforth, we deal with the sixteen gravitational potentials $\psi_{\mu \nu}$. It is convenient to decompose $\psi_{\mu \nu}$ into its symmetric and antisymmetric components, 
\begin{equation}\label{8}
 h_{\mu \nu}:=2\psi_{(\mu \nu)}, \qquad  \phi_{\mu \nu}:=2\psi_{[\mu \nu]}\,.
\end{equation}
Then,
\begin{equation}\label{9} 
g_{\mu \nu}=\eta_{\mu \nu}+h_{\mu \nu}\,,\qquad   \psi:=\eta_{\mu \nu}\psi^{\mu \nu}=\frac{1}{2}h.
\end{equation}
As in GR, it proves useful to introduce the trace-reversed potentials $\overline{h}_{\mu \nu}$,
\begin{equation}\label{10}
\overline{h}_{\mu \nu}=h_{\mu \nu}-\frac 12\eta_{\mu \nu}h\,,
\end{equation} 
so that  $\overline{h}=-h$ and 
\begin{equation}\label{11}
\psi_{\mu \nu}=\frac{1}{2}\overline{h}_{\mu \nu}+\frac 12\phi_{\mu \nu}+\frac 14\eta_{\mu \nu}h\,.
\end{equation} 

In terms of the gravitational potentials $\psi_{\mu \nu}$, the field strength is 
\begin{equation}\label{12}
C_{\mu \nu \sigma}=\partial_\mu \psi_{\sigma \nu}-\partial_\nu \psi_{\sigma \mu}
\end{equation}
and the modified field strength can be expressed as 
\begin{equation}\label{13}
\mathfrak{C}_{\mu \sigma \nu}=-\overline{h}_{\nu [\mu,\sigma]}-\eta_{\nu [\mu}\overline{h}_{\sigma ]\rho,}{}^\rho+\frac 12\phi_{\mu \sigma, \nu}+\eta_{\nu [\mu} \phi_{\sigma ] \rho,}{}^\rho\,.
\end{equation}
We also note that the Einstein tensor is given by
\begin{equation}\label{14}
G_{\mu \nu}=\partial_\sigma \mathfrak{C}_{\mu}{}^{\sigma}{}_{\nu}=-\frac
12\square\, 
\overline{h}_{\mu \nu}+\overline{h}^\rho{}_{(\mu,\nu)\rho}-\frac
12\eta_{\mu \nu}\overline{h}^{\rho \sigma}{}_{,\rho \sigma}\,,
\end{equation}
where $\square :=\eta^{\alpha \beta}\partial_\alpha \partial_\beta$.

The gravitational field equations are given in this case by Eq.~\eqref{5}, where $ \partial_\sigma{\cal H}_{\mu}{}^{\sigma}{}_{\nu}=T_{\nu \mu}$ in the general linear approximation and 
\begin{equation}\label{15}
\kappa ~ {\cal H}_{\mu}{}^{\sigma}{}_{\nu}(x)=\mathfrak{C}_{\mu}{}^{\sigma}{}_{\nu}(x)+\int{\cal K}(x, y) \mathfrak{C}_{\mu}{}^{\sigma}{}_{\nu}(y)~d^4y                 
\end{equation}
is our nonlocal ansatz. Thus the linearized gravitational field equations reduce, via Eqs.~\eqref{14} and~\eqref{15}, to the following sixteen equations for the sixteen components of $\psi_{\mu \nu}$, namely, 
\begin{equation}\label{16}
 G_{\mu \nu}(x)+\int \frac{\partial{\cal K}(x, y)}{\partial x^\sigma} \mathfrak{C}_{\mu}{}^{\sigma}{}_{\nu}(y)~d^4y= \kappa  T_{\nu \mu}(x)\,,              
\end{equation}
where $T_{\nu \mu}$ is not in general a symmetric tensor, but it is conserved, $\partial_{\mu}T^{\nu \mu}=0$, due to the fact that $\mathfrak{C}_{\mu \sigma \nu}$ is in general antisymmetric in its first two indices.

A few comments are in order here regarding our simple ansatz~\eqref{15}. Our \emph{linear} nonlocal prescription involving a \emph{scalar} kernel is by no means unique; in particular, one may contemplate a nonlinear generalization of Eq.~\eqref{15} if demanded by the confrontation of the nonlocal theory with observation. We adopt the viewpoint, developed in Ref.~\cite{19}, that the nonlocal kernel is ultimately determined via comparison with observational data. Nevertheless, it is possible to imagine that the kernel may be derivable from the field equations of a more sophisticated future theory. In any case, the conclusions of the present work are limited to our current ansatz~\eqref{15}. In connection with this ansatz, it is in general useful to introduce a constant real parameter $\gamma$, say,  in front of the integral term in Eq.~\eqref{15}. For $\gamma=0$, we recover the local theory; otherwise, $\gamma$ would parametrize nonlocal deviations from the standard local GR. 

The nature of the kernel has been discussed in some detail in Refs.~\cite{17, 18, 19, 20, 21}. In keeping with the general linear approximation under consideration here, we assume a linear kernel, in which case ${\cal K}(x, y)$ is a \emph{universal} function $K$ of $x-y$, ${\cal K}(x, y)=K(x-y)$. The causality requirement is satisfied once $K$ is nonzero only for situations where $x^\mu-y^\mu$ is a future directed timelike or null vector in Minkowski spacetime. That is, event $y$ must be within or on the past light cone of event $x$. This causality requirement implies that $K(x-y)$ is proportional to $H(x^0-y^0-|\mathbf{x}-\mathbf{y}|)$, where $H(s)$ is the Heaviside  (unit step) function such that $H(s)=1$ for $s\ge0$ and $H(s)=0$ for $s<0$. The nonlocal integral term in Eq.~\eqref{15} may thus be described as an average over past events; therefore, time-reversal invariance is in general broken in such nonlocal theories. 

The gravitational field is expected to vanish at spatial infinity. We assume that the convolution kernel vanishes for past events that are infinitely distant in space or time. Moreover, it is important to note that only the symmetric part of $\psi_{\mu \nu}$ involves propagating fields and we assume that there are no incoming gravitational waves. With a causal convolution kernel in Eq.~\eqref{16} and using $\partial K/\partial x^\sigma=-\partial K/\partial y^\sigma$, we find
\begin{equation}\label{17}
 G_{\mu \nu}(x)+\int  K(x-y)G_{\mu \nu}(y)~d^4y= \kappa  T_{\nu \mu}(x)+S_{\mu \nu}(x)\,,              
\end{equation}
where
\begin{equation}\label{18}
 S_{\mu \nu}(x)=\int \frac{\partial}{\partial y^\sigma} [K(x-y)\mathfrak{C}_{\mu}{}^{\sigma}{}_{\nu}(y)]~d^4y          
\end{equation} 
and $\kappa  T_{\nu \mu}+S_{\mu \nu}$ is a symmetric tensor. The integrand in Eq.~\eqref{18} is the sum of four terms; however, only the temporal ($\sigma=0$) integration makes a contribution to $S_{\mu \nu}(x)$ at $y^0=x^0-|\mathbf{x}-\mathbf{y}|$ and we thus find
\begin{equation}\label{19}
 S_{\mu \nu}(x)=\int K(|\mathbf{x}-\mathbf{y}|, \mathbf{x}-\mathbf{y})\mathfrak{C}_{\mu}{}^0{}_{\nu}(x^0-|\mathbf{x}-\mathbf{y}|, \mathbf{y})~d^3y\,,          
\end{equation} 
where we will henceforth refer to $|\mathbf{x}-\mathbf{y}|$ in the temporal variables as the \emph{retardation}.

The integral relationships that we consider are in general of the Fredholm type~\cite{Tr}; however, they turn into Volterra integral relations whenever the kernels are causal. Our linearized gravitational field equation is thus the integro-differential Eq.~\eqref{17}. The Liouville-Neumann approach to the solution of Eq.~\eqref{17} involves modifying this equation by taking the integral term to the right-hand side and then replacing $G_{\mu \nu} (y)$ in the integrand by its value given by the modified Eq.~\eqref{17}. Iterating this process eventually leads to an infinite Neumann series for the iterated kernels. If this series uniformly converges, we obtain a unique solution of Eq.~\eqref{17} involving a kernel $R$ that is reciprocal to $K$---see Eqs.~\eqref{32}  and~\eqref{33} below. This convergence of the Neumann series is demonstrated in Appendix A under physically reasonable conditions. As shown in Appendix A,   $K(x-y)$ has a unique \emph{reciprocal} causal convolution kernel $R(x-y)$; therefore, Eq.~\eqref{17} can be written as $G_{\mu \nu}=U_{\mu \nu}$, where
\begin{equation}\label{20}
 U_{\mu \nu}(x)=\kappa  T_{\nu \mu}(x)+S_{\mu \nu}(x) +\int  R(x-y)[\kappa  T_{\nu \mu}(y)+S_{\mu \nu}(y)]~d^4y\,.              
\end{equation}
In this equation, we think of $T_{\nu \mu}$ as the matter source and its convolution with $R$ as the effective dark matter source. The source-free equation for the amplitude of the gravitational perturbation is then given by 
\begin{equation}\label{21}
 G_{\mu \nu}(x)=S_{\mu \nu}(x) +\int  R(x-y)S_{\mu \nu}(y)~d^4y\,.              
\end{equation}

It is a significant feature of this linearized theory of nonlocal gravity that causal convolution kernels have reciprocal causal convolution kernels under physically reasonable conditions---see Appendix A. It means that nonlocality in the linearized theory can be given a consistent physical interpretation in terms of simulated (fake) dark matter source that is the convolution of the actual source of the gravitational field with the reciprocal \emph{causal} kernel.

We emphasize that in this simplest theory of nonlocal gravity, the kernels $K$ and $R$ are scalars. Therefore, once these kernels are known in the rest frame of a gravitational system, they can be determined in any other inertial system via Lorentz invariance. In the background global Minkowski spacetime, imagine a different inertial frame with inertial coordinates $x'^{\mu}$ and fundamental observers (i.e., those at rest in space) that move uniformly with respect to the background, so that $x'=\Lambda x+b$, where $\Lambda$ is a Lorentz matrix and $b$ indicates a constant spacetime translation. Then, 
\begin{equation}\label{22}
K(x-y)=K'[\Lambda(x-y)]\,, \qquad R(x-y)=R'[\Lambda(x-y)]\,.              
\end{equation}
These considerations are necessary in order to deal with the astrophysics of dark matter in moving systems, such as, for instance, the Bullet Cluster~\cite{BC1, BC2}.

Let us conclude this general treatment of linear approximation scheme by discussing the gauge freedom of the gravitational potentials. Under an infinitesimal coordinate transformation $x^\mu \mapsto x'^\mu=x^\mu-\epsilon^\mu(x)$, we find that $\psi_{\mu \nu} \mapsto \psi'_{\mu \nu}=\psi_{\mu \nu}+\epsilon_{\mu,\nu}$ to linear order in $\epsilon^\mu$. Moreover, 
\begin{equation}\label{23}
\overline{h}'_{\mu \nu}=\overline{h}_{\mu \nu}+\epsilon_{\mu,\nu}+\epsilon_{\nu,\mu}-\eta_{\mu \nu}\epsilon^\alpha{}_{,\alpha}\,, \qquad   \phi'_{\mu \nu}=\phi_{\mu \nu}+\epsilon_{\mu,\nu}-\epsilon_{\nu,\mu}
\end{equation} 
and we note in passing that 
\begin{equation}\label{24}
\overline{h}'=\overline{h}-2\epsilon^\alpha{}_{,\alpha}\,. 
\end{equation} 
As expected from the electrodynamic analogy, the field strength $C_{\mu \nu \sigma}$ is invariant under a gauge transformation; furthermore, the same holds for the modified field strength $\mathfrak{C}_{\mu \sigma \nu}$. Thus Eqs.~\eqref{14}--\eqref{21} are all gauge invariant. Further discussion of the gauge freedom of gravitational potentials is contained in Appendix B.

We now consider the physical consequences of the linear approximation scheme, namely, the correspondence with Newtonian gravity and linearized gravitational waves. 

\section{Nonlocal Newtonian Gravity}

The correspondence of nonlocal gravity with the Newtonian theory of gravitation can be established in much the same way as in GR. Indeed, we assume that the gravitational potentials $\psi_{\mu \nu}$ form a \emph{diagonal} matrix that depends only on the Newtonian gravitational potential $\Phi$. The resulting modified Poisson's equation for $\Phi$ is independent of the speed of light $c$; therefore, one may think of the correspondence with  Newtonian gravity as a formal transition in which $c \to \infty$.

We start with Eq.~\eqref{16} and impose the transverse gauge condition $\overline{h}^{\mu \nu}{}_{,\nu}=0$ (see Appendix B).  Assuming that the only nonzero component of $\overline{h}_{\mu \nu}$ that is relevant here is $\overline{h}_{0 0}=-4\Phi/c^2$, we find that $2\psi_{\mu \nu}=h_{\mu \nu}$ is given by $(-2\Phi/c^2)$~diag$(1,1,1,1)$. The only significant component of Eq.~\eqref{16} is thus the $\mu=\nu=0$ one with $c^2\mathfrak{C}_{0i0}=2\Phi_{,i}$ and $T_{00}=\rho c^2$, where $\rho$ is the matter density. Moreover, we assume that in the Newtonian limit, 
\begin{eqnarray}\label{25}
K(x-y)=\delta(x^0-y^0)\chi(\mathbf{x}-\mathbf{y})\,,
\end{eqnarray}
due to the absence of any retardation effects for $c \to \infty$. The reciprocal kernel is then of a similar form~\cite{19, 21}
\begin{eqnarray}\label{26}
R(x-y)=\delta(x^0-y^0)q(\mathbf{x}-\mathbf{y})\,
\end{eqnarray}
and Eq.~\eqref{16} reduces in this case to the interesting form
\begin{equation}\label{27}
  \nabla^2\Phi (\mathbf{x}) = 4\pi G\Big[\rho(\mathbf{x})+  \int q(\mathbf{x}-\mathbf{y}) \rho(\mathbf{y})d^3y\Big]\,.
\end{equation}

This nonlocal modification of Poisson's equation has been examined in detail in Ref.~\cite{21}. Nonlocality appears in Eq.~\eqref{27} as an extra (``dark") matter source whose density is the convolution of matter density with the reciprocal Newtonian kernel. That is, nonlocality simulates dark matter~\cite{RF, RW, SR}. It is remarkable that, within the phenomenological Tohline-Kuhn approach to the problem of rotation curves of spiral galaxies~\cite{T, K, B}, an equation of the form~\eqref{27} was suggested by Kuhn in order to account for dark matter as modified gravity. In fact, Tohline~\cite{T} first suggested that the Newtonian gravitational potential for a point mass $M$ be replaced by
\begin{equation}\label{28}
\Phi(\mathbf{x})=-\frac{GM}{|\mathbf{x}|}
+\frac{GM}{\lambda_0}\ln\left(\frac{|\mathbf{x}|}{\lambda_0}\right)\,,
\end{equation}
where $\lambda_0$ is a constant length of order 1\,kpc. Tohline's purely phenomenological proposal was later extended by Kuhn and his collaborators in order to resolve the problem of dark matter in galaxies and clusters of galaxies~\cite{K}. In particular, Kuhn suggested a modification of Poisson's equation of the form~\eqref{27} with the kernel 
\begin{equation}\label{29}
Q(\mathbf{x}-\mathbf{y})=\frac{1}{4\pi\lambda_0}
\frac{1}{|\mathbf{x}-\mathbf{y}|^2}\,.
\end{equation}
This Kuhn kernel $Q$ is such that for $\rho(\mathbf{x})=M \delta(\mathbf{x})$, Eq.~\eqref{28} is a solution of the modified Poisson equation~\eqref{27}. Further implications of the Tohline-Kuhn approach are discussed in Bekenstein's lucid review article~\cite{B} and, in connection with nonlocal gravity, in Refs.~\cite{17, 18, 19, 20, 21}. In particular, we should mention here that the Tohline-Kuhn scheme is in seeming conflict with the Tully-Fisher relation~\cite{TF}. As discussed in Ref.~\cite{21}, we take the view that a purely gravitational treatment does not contain enough physics to deal with electromagnetic radiation aspects needed for a fair comparison of the implications of nonlocal gravity with the empirical Tully-Fisher law~\cite{TF}. 

To determine the kernel of nonlocal general relativity from comparison with observational data, it is necessary to extend the Kuhn kernel over all space. This issue has been extensively treated in Ref.~\cite{21}; in particular, two examples were completely worked out explicitly. These can be expressed via the reciprocal kernel $q(\mathbf{x}-\mathbf{y})$ as
\begin{equation}\label{30}
q_1=\frac{1}{4\pi \lambda_0}~ \frac{1+\alpha(a+u)}{(a+u)^2}~e^{-\alpha u}\,,
\end{equation}
\begin{equation}\label{31}
q_2=\frac{1}{4\pi \lambda_0}~ \frac{1+\alpha(a+u)}{u(a+u)}~e^{-\alpha u}\,,
\end{equation}
where $u=|\mathbf{x}-\mathbf{y}|$ and $\lambda_0=10~$kpc. Typical values of the parameters $\alpha$ and $a$ are $\alpha^{-1}=10~ \lambda_0$ and $a=10^{-3} \lambda_0$, so that in general $\alpha \lambda_0$ and $a/\lambda_0$ are positive constants that are small compared to unity and $0<\alpha a \ll1$. We note that $q_1$ and $q_2$ are smooth positive functions that rapidly fall off to zero at infinity; moreover, they are integrable as well as square integrable. The explicit forms of the corresponding $\chi$ kernels have been numerically determined and presented in Ref.~\cite{21}. It follows from the numerical results that $\chi_1$ and $\chi_2$ drop off with distance $u$ extremely fast and are essentially zero beyond around $2.5~\lambda_0$. To deal theoretically with isolated astronomical systems, one routinely assumes that they possess sharp boundaries and are therefore compactly supported. In a similar way, we expect that in practice the dark counterpart of an isolated astronomical system is more extended but still can be cut off beyond a certain distance away from the source and hence considered isolated as well.

The novel features of nonlocal gravity emerge on galactic scales. That is, the deviation of the nonlocal theory from general relativity is associated with a length scale of order $\lambda_0$. Indeed, nonlocality disappears for $\lambda_0 \to \infty$. Newton's theory of gravitation and its relativistic extension in classical GR are devoid of any intrinsic length scale~\cite{Me}; moreover, GR is in good agreement with current solar system data. Whether this situation extends to galactic scales is an open problem, however. We assume that what appears as dark matter in astrophysics is in fact mainly a manifestation of the nonlocal aspect of the gravitational interaction. 

Turning now to the problem of gravitational radiation in nonlocal general relativity, the first task before us is the determination of the reciprocal kernel $R(x-y)$ and hence the kernel of the theory $K(x-y)$. This is treated in the next section.  

\section{Kernels}

In the linear weak-field approximation, we have neglected possible nonlinearities in kernel ${\cal K}$ and have assumed that our convolution kernel $K(x-y)$ is such that it has a reciprocal $R(x-y)$. In the Newtonian regime, it is the reciprocal kernel that can be \emph{directly} compared with observational data. Following our general treatment in Ref.~\cite{21}, the relation between $K$ and $R$ may be expressed---via Fredholm integral equations of the second kind~\cite{Tr}---as
\begin{equation}\label{32}
{\cal G}(x)+ \int K(x-y) {\cal G}(y)d^4y = {\cal F}(x)\,,
\end{equation}
\begin{equation}\label{33}
{\cal F}(x)+ \int R(x-y) {\cal F}(y)d^4y = {\cal G}(x)\,.
\end{equation}
The Liouville-Neumann method of successive substitutions makes it formally possible to relate these equations~\cite{Tr}. However, we found by way of several trials in Ref.~\cite{21} that in the Newtonian regime, where the kernels are given by Eqs.~\eqref{25} and~\eqref{26}, the corresponding Neumann series would converge only in situations that were outside the physical domain of interest and so we had to resort to the Fourier transform method. 

Beyond the Newtonian limit, as explained in detail in Appendix A, the causality requirement renders $K$ and $R$ Volterra kernels and Eqs.~\eqref{32} and~\eqref{33} Volterra integral  equations of the second kind. The imposition of causality makes it possible to show, via the Liouville-Neumann method applied to the Volterra algebra that a unique reciprocal kernel exists and is causal~\cite{MR, FV}---see Appendix A. Restricting our treatment to continuous, absolutely integrable and square integrable causal kernels, it is in principle also possible to apply the Fourier transform method.   

Let
\begin{equation}\label{34}
\hat{f} (\xi) =  \int f(x) e^{-i \xi \cdot x}~ d^4x\,
\end{equation}
be the Fourier transform of $f$, where $\xi \cdot x := \eta_{\alpha \beta}\xi^\alpha x^\beta$. Then, 
\begin{equation}\label{35}
f(x) = \frac{1}{(2\pi)^4} \int \hat{f} (\xi) e^{i \xi \cdot x}~ d^4\xi\,.
\end{equation}
As described in detail in Ref.~\cite{21}, we work in the space of functions for which such operations are permissible. It follows from Eqs.~\eqref{32} and~\eqref{33} via Fourier transformation that 
\begin{equation}\label{36}
\hat{K} (\xi) =- \frac{\hat{R} (\xi)}{1+\hat{R} (\xi)}\,, \qquad \hat{R} (\xi) =- \frac{\hat{K} (\xi)}{1+\hat{K} (\xi)}\,,
\end{equation}
since the kernels are reciprocal. Starting with kernel $R$, one can find $K$ provided $1+\hat{R} (\xi)\ne 0$, etc. In Ref.~\cite{21}, working in the Newtonian regime ($c\to \infty$), we showed that for an appropriate reciprocal kernel $q$, which would be a natural extension of the Kuhn kernel $Q$ to all space, kernel $\chi$ exists with 
\begin{equation}\label{37}
\hat {\chi} (|\boldsymbol{\xi}|)=-\frac{\hat{q} (|\boldsymbol{\xi}|)}{1+\hat{q} (|\boldsymbol{\xi}|)}\,,
\end{equation}
provided $1+\hat{q} (|\boldsymbol{\xi}|)\ne 0$. In particular, for the class of kernels of either form $q_1$ given by Eq.~\eqref{30} or $q_2$ given by Eq.~\eqref{31}, we showed that~\cite{21}
\begin{equation}\label{38}
1+\hat{q} (|\boldsymbol{\xi}|)>0\,.
\end{equation}

To proceed further, we need to find the functional form of $R$ and $K$ in order to go beyond the Newtonian limit ($c=\infty$). Therefore, in what follows we wish to concentrate on the Dirac delta function that appears in the Newtonian kernels~\eqref{25} and~\eqref{26}. In general, we expect that \emph{constitutive} kernels $R$ and $K$ should decay exponentially for events that are distant in space \emph{and time}. The spatial exponential decay is already evident in Eqs.~\eqref{30} and~\eqref{31}. To see how a corresponding \emph{temporal} decay could come about, we recall here a mathematical result, namely, that for $n=1, 2, 3,...,$ the functions
\begin{equation}\label{39}
\delta_n(s)=ne^{-ns}H(s)\,
\end{equation}
form a Dirac sequence such that as $n\to \infty$, $\delta_n(s) \to \delta(s-0^+)$. It follows that for a positive constant length $1/A$, we have that as $c\to \infty$,
\begin{equation}\label{40}
H(t_x-t_y-\frac{1}{c}|\mathbf{x}-\mathbf{y}|)A~c~e^{-Ac(t_x-t_y)} \to \delta(t_x-t_y)\,,
\end{equation}
where the singularity in $\delta(t_x-t_y)$ is at $t_x-t_y=0^+$. Here the role of $n$ is formally played by a fixed positive constant times $c$. The retardation in Eq.~\eqref{40} is proportional to $1/c$ and tends to zero as $c$ formally approaches $\infty$. Moreover, we note that $\delta(t_x)=c~\delta(x^0)$. Based on these considerations, we will henceforth assume that 
\begin{equation}\label{41}
R(x-y)=H(x^0-y^0-|\mathbf{x}-\mathbf{y}|)A~e^{-A(x^0-y^0)} q(\mathbf{x}-\mathbf{y})\,,
\end{equation}
where $1/A$ is a constant length of order $\lambda_0$; in fact, for the estimates in this paper we set $A=\alpha=1/(10~ \lambda_0)$. Clearly in the limit as $c\to \infty$, we recover Eq.~\eqref{26}. Moreover, we assume that $q(\mathbf{x}-\mathbf{y})$ is of the type of $q_1$ or $q_2$ for which Eq.~\eqref{38} is satisfied. It follows from these considerations that $R(t_x, \mathbf{x})$ is a positive function on as well as within the future light cone and zero otherwise; moreover, it is integrable as well as square integrable, and rapidly decreases to zero at infinity.

Apropos of the functional form of our scalar convolution kernel~\eqref{41}, it is interesting to digress here and mention a solution $\Psi$ of the massless scalar wave equation, $\square \Psi=0$, such that $\Psi(ct, \mathbf{x})=P(r)\exp{(-\alpha c t)}$ with $r=|\mathbf{x}|$ and constant $\alpha>0$. Then, $\nabla^2P=\alpha^2 P$ and $rP(r)$ is a linear combination of $\exp{(-\alpha r)}$ and $\exp{(\alpha r)}$. Thus a possible solution for $\Psi$ is of the form $r^{-1}\exp{[-\alpha(ct+r)]}$, which is in some ways reminiscent of Eq.~\eqref{41}.

The kernel given in Eq.~\eqref{41} is causal by construction; its reciprocal, namely, $K$, is causal as well. The proof of this assertion in a rather general context using the Liouville-Neumann method is essentially due to M.~Riesz~\cite{MR, FV} and is briefly described in Appendix A. The corresponding proof using the general Fourier transform method appears to be rather complicated. However, to demonstrate the consistency of our approach in a simple mathematical setting that employs only dominant terms as $c$ tends to infinity, let us neglect retardation effects and write Eq.~\eqref{41} instead in the form
\begin{equation}\label{42}
R(x-y)\sim H(x^0-y^0)A~e^{-A(x^0-y^0)} q(\mathbf{x}-\mathbf{y})\,.
\end{equation}
It then follows from Fourier transforming Eq.~\eqref{42} that 
\begin{equation}\label{43}
\hat{R}(\xi) \sim \frac{A}{A-i\xi^0}~\hat{q} (|\boldsymbol{\xi}|)\,,
\end{equation}
which satisfies the requirement that $1+\hat{R}\ne0$. Therefore, we can find $\hat{K}$ using Eq.~\eqref{36},
\begin{equation}\label{44}
\hat{K} (\xi) \sim - \frac{A~ \hat{q} (|\boldsymbol{\xi}|)}{-i\xi^0+A~[1+\hat{q} (|\boldsymbol{\xi}|)]}\,.
\end{equation}
In calculating the Fourier transform of Eq.~\eqref{44}, it is useful to note that one can use contour integration and Jordan's lemma to find
\begin{equation}\label{45}
\int_{-\infty}^{\infty}\frac{e^{-i\xi^0 x^0}}{-i\xi^0+A~(1+\hat{q})}~d\xi^0=2\pi H(x^0)e^{-A(1+\hat{q})x^0}\,.
\end{equation}
Here, in the complex $\xi^0$ plane, there is just a simple pole singularity at $-iA(1+\hat{q})$, which occurs in the lower half-plane due to the fact that $A(1+\hat{q})>0$ by assumption---cf. Eq.~\eqref{38}. Therefore, 
\begin{equation}\label{46}
K(x) \sim -\frac{A}{(2\pi)^3}H(x^0) \int \hat{q} (|\boldsymbol{\xi}|) e^{i \boldsymbol{\xi} \cdot \mathbf{x}}e^{-A(1+\hat{q})x^0}~ d^3\xi\,.
\end{equation}
Let us recall that in the limit as $c\to \infty$, we have
\begin{equation}\label{47}
H(x^0)A(1+\hat{q})e^{-A(1+\hat{q})x^0} \to \delta(x^0)\,.
\end{equation}
Moreover, it follows from Eq.~\eqref{37} that
\begin{equation}\label{48}
\frac{1}{(2\pi)^3} \int \Big[-\frac{\hat{q} (|\boldsymbol{\xi}|)}{1+\hat{q} (|\boldsymbol{\xi}|)}\Big] e^{i \boldsymbol{\xi} \cdot \mathbf{x}}~ d^3\xi=\chi (\mathbf{x})\,.
\end{equation}
Putting Eqs.~\eqref{46}--\eqref{48} together, we find that, in agreement with Eq.~\eqref{25}, $K(x)=\delta(x^0)\chi(\mathbf{x})$ in the Newtonian limit, as expected. 

The reciprocity between $K$ and $R$ indicates that our starting point, Eq.~\eqref{41}, is far from unique; for instance, Eq.~\eqref{46} has essentially the same properties as Eq.~\eqref{42}. However, we will work with Eqs.~\eqref{41} and~\eqref{42} in what follows for the sake of simplicity. Furthermore, neglecting retardation has led to a manageable expression for the kernel $K$, namely, Eq.~\eqref{46}, which will be employed in the following sections. The nature of this simplification is examined in Appendix C.

\section{Gravitational Waves: Propagation}

Let us return to the linearized nonlocal field equations and impose the transverse gauge condition $\overline{h}^{\mu \nu}{}_{,\nu}=0$. As is well known from GR, this condition simplifies the linearized Einstein tensor such that $-2G_{\mu \nu}=\square\, \overline{h}_{\mu \nu}$. The linearized field equations can now be written as
\begin{equation}\label{49}
\square\, \overline{h}_{\mu \nu}=-2U_{\mu \nu}(x)\,,
\end{equation}
where $U_{\mu \nu}$ is given by Eq.~\eqref{20}. 

Assuming that the source $T_{\mu \nu}$ is isolated, the corresponding dark source is also then expected to be in effect isolated due to the rapid spatial decay of the reciprocal kernel. Far  from the source and its dark counterpart, the gravitational potentials in the wave zone satisfy the field equations
\begin{equation}\label{50}
\square\, \overline{h}_{\mu \nu}+2S_{\mu \nu}(x)+2\int R(x-y)S_{\mu \nu}(y)d^4y=0\,.
\end{equation}
Turning now to the expression for $S_{\mu \nu}$ given in Eq.~\eqref{19}, we observe that $\mathfrak{C}_{00\nu}=0$ due to the antisymmetry of the modified field strength in its first two indices; therefore, $S_{0 \nu}=0$, but $S_{\mu 0}$ is in general nonzero. It follows from Eq.~\eqref{50} that $S_{0 \nu}=0$ implies that $\square\, \overline{h}_{0 \nu}=0$. Far in the wave zone, a fixed detector perceives the emitted gravitational radiation potentials $\overline{h}_{0 \nu}$ to be essentially  \emph{plane} gravitational waves. Therefore, following the same analysis as in standard GR, given here in Appendix B for the sake of completeness, one can choose the remaining gauge degrees of freedom to set $\overline{h}_{0 \nu}=0$ and $\overline{h}=0$. Thus with a suitable choice of gauge, $h_{\mu \nu}$ is purely spatial and traceless with $h^{ij}{}_{,j}=0$.

Furthermore, it follows from Eq.~\eqref{50} that 
\begin{equation}\label{51}
S_{i0}(x)+\int R(x-y)S_{i0}(y)d^4y=0\,,
\end{equation}
which implies, via Fourier transformation, that $S_{i0}(x)=0$, since $1+\hat{R} (\xi)\ne 0$. Thus 
\begin{equation}\label{52}
 S_{i0}(x)=\int K(|\mathbf{x}-\mathbf{y}|, \mathbf{x}-\mathbf{y})\mathfrak{C}_{i}{}^0{}_{0}(x^0-|\mathbf{x}-\mathbf{y}|, \mathbf{y})~d^3y=0\,,          
\end{equation} 
where
\begin{equation}\label{53}
\mathfrak{C}_{i}{}^0{}_{0}=-\frac{1}{2}~\phi_{ij,}{}^j\,.         
\end{equation} 
In the decomposition of the gravitational potentials $\psi_{\mu \nu}$ into symmetric and antisymmetric parts, only the symmetric part propagates and the antisymmetric part satisfies Eqs.~\eqref{52}--\eqref{53}; therefore, to simplify matters, we shall set 
\begin{equation}\label{54}
\phi_{\mu \nu}=0\,      
\end{equation} 
for the rest of this paper. 

Summing up, in the wave zone the gravitational potentials reduce to $h_{ij}(x)$ such that 
\begin{equation}\label{55}
\square\, h_{ij}+2S_{ij}(x)+2\int R(x-y)S_{ij}(y)d^4y=0\,,
\end{equation}
where in expression~\eqref{19} for $S_{ij}(x)$,
\begin{equation}\label{56}
\mathfrak{C}_{i}{}^0{}_{j}=\frac{1}{2}~\frac{\partial h_{ij}}{\partial t}\,,         
\end{equation} 
since now in Eq.~\eqref{13} the transverse gauge condition holds; moreover, $\overline{h}_{0\mu}=0$, $\overline{h}=0$ and $\phi_{\mu \nu}=0$.

To solve Eq.~\eqref{55}, we work in the Fourier domain, where this equation can be expressed as 
\begin{equation}\label{57}
(\omega^2-|\mathbf{k}|^2)\hat{h}_{ij}(\omega, \mathbf{k})+2(1+\hat{R})\hat{S}_{ij}=0\,.
\end{equation}
Here $\omega$ is the wave frequency and $\mathbf{k}$ is the wave vector. These form the components of a point in the 4D Fourier domain characterized by the propagation vector $k^\mu=(\omega/c, \mathbf{k})$. Using the approach adopted in section IV, where retardation effects are neglected, $\hat{S}_{ij}$ can be calculated from Eqs.~\eqref{19},~\eqref{46} and~\eqref{56}, and the result is
\begin{equation}\label{58}
2 \hat{S}_{ij}\sim iA\omega \hat{q}(|\mathbf{k}|) \hat{h}_{ij}(\omega, \mathbf{k})\,.
\end{equation}
Moreover, $\hat{R}$ is given by Eq.~\eqref{43}. From these results, the dispersion relation for gravitational waves takes the approximate form
\begin{equation}\label{59}
\omega^2- |\mathbf{k}|^2 \sim -iA\omega (1+\hat{R}) \hat{q}(|\mathbf{k}|)\,.
\end{equation}
The complex nature of this relation indicates the possible presence of decaying or growing modes; however, it follows from Eq.~\eqref{43} that in our approximation scheme all of the modes  indeed decay. 
A complete analysis of the solutions of the nonlocal wave Eq.~\eqref{55} is beyond the scope of this work. We show in the rest of this section that the nonlocal contribution to the dispersion relation~\eqref{59} is actually negligible for current efforts to detect gravitational radiation.  

The scale associated with nonlocality, $\lambda_0=10$ kpc, corresponds to a characteristic frequency $\nu_0=c/\lambda_0 \approx 10^{-12}$ Hz. We expect that the propagation of such low-frequency gravitational waves would be significantly affected by nonlocality. However, current laboratory efforts~\cite{Ri} are directed at detecting gravitational waves of dominant frequency $\gtrsim 1$ Hz and corresponding dominant wavelength $\lambda$, where $\lambda / \lambda_0 \lesssim 10^{-12}$. Moreover, future space-based interferometers may be able to detect low-frequency ($\sim 10^{-4}$ Hz) gravitational waves from astrophysical sources~\cite{Ri}. On the other hand, pulsar timing residuals from an ensemble of highly stable pulsars can be used to search for a stochastic background of very low-frequency ($\sim$ several nHz) gravitational waves~\cite{Ri}. Thus current observational possibilities all involve gravitational radiation of wavelengths that are short compared with $\lambda_0$~\cite{Ri}; indeed, in all cases of interest, 
\begin{equation}\label{60}
\frac{\lambda}{\lambda_0} \lesssim 10^{-4}\,.
\end{equation}

For such wavelengths the contribution of nonlocality to the dispersion relation~\eqref{59} is negligibly small, so that in practice $\omega^2 \approx |\mathbf{k}|^2$. To see this, we first note that Eq.~\eqref{43} implies
\begin{equation}\label{61}
|\hat{R}(\omega, \mathbf{k})| \sim \frac{A}{(A^2+\omega^2)^{1/2}}~|\hat{q} (|\mathbf{k}|)|\,,
\end{equation}
where only positive square roots are considered in this paper. Moreover, to simplify matters still further, we can estimate $|\hat{q} (|\mathbf{k}|)|$ using the Kuhn kernel instead. For the Kuhn kernel $Q$ given by Eq.~\eqref{29}, we have that 
\begin{equation}\label{62}
\hat{Q}(|\mathbf{k}|)=\frac{\pi}{2\lambda_0} \frac{1}{|\mathbf{k}|}\,.
\end{equation}
Thus with $A=\alpha=1/(10~ \lambda_0)$, $A/\omega=(20\pi)^{-1}\lambda/\lambda_0$; hence, we find $|\hat{R}|\sim (80\pi)^{-1}(\lambda/\lambda_0)^2$, which is negligible compared to unity. Similarly, the ratio of the nonlocal term in the dispersion relation~\eqref{59} to $\omega^2$ has the absolute magnitude $\approx A\omega^{-1} |\hat{q}|$, which is again of order $(80 \pi)^{-1}(\lambda/\lambda_0)^2$; hence, the nonlocal contribution to Eq.~\eqref{59} can be neglected. 

Unless observational efforts extend to extremely low frequencies of order $10^{-12}$ Hz, we can safely neglect nonlocal terms in Eq.~\eqref{55}, so that $\square\, h_{ij}\approx 0$. Therefore, for current observational possibilities, gravitational wave propagation in the nonlocal theory reduces essentially to that of GR and we can recover the TT gauge and hence the two independent helicity states of gravitational radiation---see Appendix B. 

We now turn to the influence of the dark matter source upon the amplitude of emitted gravitational radiation. 

\section{Gravitational Waves: Dark Source}

The general linear approximation of section II is such that $\partial_\mu G^{\mu \nu}=0$ in Eq.~\eqref{14} and $\partial_\mu T^{\nu \mu}=0$ in Eq.~\eqref{16}. Based on the considerations of section V, we can now essentially ignore $S_{\mu \nu}$ in Eqs.~\eqref{17} and~\eqref{20}. This implies that in the linearized field equations given by Eq.~\eqref{49}, we have $U_{\mu \nu}\approx \kappa {\cal T}_{\mu \nu}$, where $ {\cal T}_{\mu \nu}$ consists of the contribution of the source and the dark source, namely, 
\begin{equation}\label{D1}
{\cal T}^{\mu \nu}:=T^{\mu \nu}+T_D^{\mu \nu}\,.
\end{equation}
Here, the dark source is represented by
\begin{equation}\label{D2}
T_D^{\mu \nu}(x)=\int R(x-y)T^{\mu \nu}(y)d^4y\,
\end{equation}
and both energy-momentum tensors are now symmetric and independently conserved in the approximation scheme under consideration here. We will treat the source and its dark counterpart as isolated astronomical systems.

We are interested in the special retarded solution of the linearized field equations given by
\begin{equation}\label{D3}
 \overline{h}^{\mu \nu}(x^0, \mathbf{x}) \approx \frac{\kappa}{2\pi}\int \frac{{\cal T}^{\mu \nu}(x^0-|\mathbf{x}-\mathbf{y}|, \mathbf{y})}{|\mathbf{x}-\mathbf{y}|}~d^3y\,.         
\end{equation} 
Far away from the source, we can introduce in Eq.~\eqref{D3} the approximation that $|\mathbf{x}-\mathbf{y} |\approx  |\mathbf{x}|=r$; that is,
\begin{equation}\label{D4}
 \overline{h}^{\mu \nu}(x^0, \mathbf{x}) \approx \frac{\kappa}{2\pi r}\int {\cal T}^{\mu \nu}(x^0-r, \mathbf{y})~d^3y\,,         
\end{equation} 
so that the solution takes the form of a spherical wave approaching the detector. Furthermore, let $\mathbf{n}$ denote the unit vector that represents the direction of propagation from the source to the receiver. Far in the wave zone, the spherical wave front can be \emph{locally} approximated by a plane wave front with wave vector $\mathbf{k}=\omega \mathbf{n}$. The wave amplitude in the TT gauge---see Appendix B---can be extracted from Eq.~\eqref{D4} by means of the projection operator $P^i{}_j=\delta^i_j-n^in_j$, namely,
\begin{equation}\label{D5}
h_{TT}^{ij}=(P^i{}_\ell P^j{}_m -\frac 12 P^{ij}P_{\ell m})\overline{h}^{\ell m}\,.       
\end{equation} 
For instance, if the spatial frame is oriented such that $\mathbf{n}$ points in the positive $x^3$ direction, the only nonzero components of $h_{TT}^{ij}$ are $h_{TT}^{11}=-h_{TT}^{22}=(\overline{h}^{11}-\overline{h}^{22})/2$ and $h_{TT}^{12}=h_{TT}^{21}=\overline{h}^{12}$.

Next, we recall that the conserved symmetric energy-momentum tensor ${\cal T}^{\mu \nu}$ of an isolated system satisfies Laue's theorem, namely, 
\begin{equation}\label{D6}
\int{\cal T}^{ij}(ct, \mathbf{x})~d^3x=\frac{1}{2c^2}\frac{d^2}{dt^2} \int {\cal T}^{00}(ct, \mathbf{x})x^ix^j~d^3x\,.
\end{equation}
It follows from Eqs.~\eqref{D4}--\eqref{D6} that, among other things, $h_{TT}^{ij}$ will depend upon the second temporal derivative of the total quadrupole moment of the system. We note here for the sake of completeness that the temporal coordinate of the quadrupole moment is in fact the ``retarded" time $ct-r$, where $r$ can be treated as a constant for the purposes of the present discussion. Moreover, the quadrupole moment could just as well be replaced by the reduced (i.e., traceless) quadrupole moment in the expression for $h_{TT}^{ij}$. 

We are particularly interested here in the contribution of the dark quadrupole moment 
\begin{equation}\label{D7}
 {\cal Q}_D^{ij}(x^0):=\int T_D^{00}(x^0, \mathbf{x})x^ix^j~d^3x 
  \end{equation}
to the total quadrupole moment of the system. It follows from the definition of dark energy density that 
\begin{equation}\label{D8}
 {\cal Q}_D^{ij}(x^0)
    = \int d^4y~T^{00}(y) \int R(x^0-y^0, \mathbf{x}-\mathbf{y})x^ix^j~d^3x\,.
\end{equation}
Let us introduce the new variable $\mathbf{u}= \mathbf{x}-\mathbf{y}$ in Eq.~\eqref{D8}; then, using expression~\eqref{42} for the reciprocal kernel and recalling that $q(\mathbf{u})$ is assumed to be spherically symmetric and of the general form of either $q_1$ or $q_2$, we find
\begin{eqnarray}\label{D9}
 {\cal Q}_D^{ij}(x^0)\sim Mq^{ij} + \frac{M_D}{M}\Sigma^{ij}(x^0)\,,
 \end{eqnarray}
where
\begin{equation}\label{D10}
M=\int T^{00}(y)~d^3y\,,  \qquad \frac{M_D}{M}=\int q(\mathbf{u})~d^3u  \, 
\end{equation} 
and 
\begin{equation}\label{D11}
q^{ij}:=\int q(\mathbf{u})u^iu^j~d^3u\,. 
\end{equation} 
Here $M$ is the mass-energy of the radiating system, which is conserved at the linear order, and $M_D$ is a rough estimate for the corresponding ``dark" mass-energy. That is, $M_D$ would be the dark mass-energy for a \emph{pointlike} source of mass-energy $M$; in fact, $M_D/M\approx 2/(\alpha \lambda_0)$ for either $q_1$ or $q_2$ with $\alpha \lambda_0=0.1$---see Ref.~\cite{21}. The time-dependent part of the ``dark" quadrupole moment may be written as  
\begin{equation}\label{D12}
\Sigma^{ij}(x^0)=A\int H(x^0-y^0)e^{-A(x^0-y^0)}{\cal Q}^{ij}(y^0)~dy^0\,, 
\end{equation} 
which is a certain average of the quadrupole moment ${\cal Q}^{ij}(t)$ of the system.

Suppose we are interested in an astronomical system whose quadrupole moment varies with time with a dominant frequency $\Omega$ that can be detected on Earth via a gravitational wave detector within a reasonable span of time; therefore, we expect that $\Omega\gg A$. Expressing such a Fourier component of ${\cal Q}^{ij}(y^0)$ as a constant amplitude times $\cos{(\Omega y^0+\varphi)}$, where $\varphi$ is a constant phase, it follows from 
\begin{equation}\label{D13}
A\int H(x^0-y^0)e^{-A(x^0-y^0)}\cos{(\Omega y^0+\varphi)}~dy^0= \frac{A}{A^2+\Omega^2}~[A \cos{(\Omega x^0+\varphi)}+\Omega \sin{(\Omega x^0+\varphi)}]\, 
\end{equation} 
that the relative contribution of the time-dependent part of the dark quadrupole moment will be reduced at least by a factor of $A/\Omega \ll 1$. We therefore conclude that the contribution of the dark source to $h_{TT}^{ij}$ is essentially negligible for all systems that are currently under consideration as possible sources of gravitational radiation that could be detectable in the near future.

\section{Gravitational Radiation Flux}

Finally, we must compute the flux of gravitational radiation energy at the detector. The energy-momentum tensor of gravitational waves ${\cal E}_{\mu \nu}$ can be calculated in the linear approximation using Eq.~\eqref{6}; that is, ${\cal E}_{\mu \nu}$ is the linearized form of the energy-momentum tensor of the gravitational field. In this calculation, we assume, as before, that $\phi_{\mu \nu}=0$; moreover, of the various gauge conditions leading to the TT gauge, at this point only $\overline{h}^{\mu \nu}{}_{, \nu}=0$ and $\overline{h}=0$ are explicitly imposed here for the sake of simplicity. Hence, we find that
\begin{equation}\label{F1}
C_{\mu \nu \sigma}=\mathfrak{C}_{\mu \nu  \sigma }=\frac 12(h_{\sigma \nu , \mu} -h_{\sigma \mu , \nu})\,
\end{equation}
and 
\begin{eqnarray}\label{F2}
 \kappa~ {\cal E}_{\mu \nu}=-\frac 14 \eta_{\mu \nu}~C^{\alpha \beta \gamma}(x)\Big[C_{\alpha \beta \gamma}(x)+\int K(x-y)C_{\alpha \beta \gamma}(y)~d^4y\Big] \nonumber  \\  
+ C_{\mu}{}^{\alpha \beta}(x)\Big[C_{\nu \alpha \beta}(x)+\int K(x-y)C_{\nu \alpha \beta}(y)~d^4y\Big]\,.      
\end{eqnarray}      
Let us note that ${\cal E}_{\mu \nu}$ is traceless and that in the absence of nonlocality (i.e., $K=0$), it becomes symmetric as well, and its form is then reminiscent of the electromagnetic energy-momentum tensor, as expected from the linearized GR$_{||}$ theory. This analogy with electrodynamics regarding the field energy and momentum has been discussed in a general context in Ref.~\cite{OPR}; moreover, it has been shown there that in GR$_{||}$, the energy-momentum tensor vanishes for a class of \emph{exact} plane-fronted gravitational waves. These exact solutions are therefore physically meaningless. This lack of physical significance of such \emph{exact plane waves} has no bearing on our \emph{local plane waves}, as we deal with the circumstance that, far from the source, the spherical gravitational waves can be approximated locally---i.e., near the receiver---by plane waves. 

We determine the energy flux implied by Eq.~\eqref{F2} in two steps. First, we compare the local ($K=0$) result ${\cal E}^{(0)}_{\mu \nu}$ to the Landau-Lifshitz energy-momentum tensor $t_{\mu \nu}$ of gravitational waves in the linear approximation of GR. Then, we give an estimate of the nonlocal contribution to ${\cal E}_{\mu \nu}$. 

Under the same explicit gauge conditions for the deviation of the metric tensor from the Minkowski metric tensor, namely, $h=0$ and $h^{\mu \nu}{}_{, \nu}=0$, the corresponding Landau-Lifshitz tensor $t_{\mu \nu}$, which is in general symmetric, but not traceless, is given by (see the Appendix of  Ref.~\cite{Mn})
\begin{eqnarray}\label{F3}
 \kappa~t_{\mu \nu}=\frac 12 h_{\mu \alpha,\beta}h_{\nu}{}^{\alpha}{}_{,}{}^{\beta}+\frac 14 \Big(h_{\alpha  \beta,\mu}h^{\alpha \beta}{}_{,\nu}-\frac 12 \eta_{\mu \nu}~h_{ \alpha \beta,\gamma}h^{\alpha \beta}{}_{,}{}^{\gamma}\Big) \nonumber \\
-\frac 12 \Big(h_{\mu \alpha,\beta}h^{\alpha \beta}{}_{,\nu}+h_{\nu \alpha,\beta}h^{\alpha \beta}{}_{,\mu}-\frac 12 \eta_{\mu \nu}~h_{ \alpha \beta,\gamma}h^{\gamma \alpha}{}_{,}{}^{\beta}\Big)\,.
\end{eqnarray}      
It is straightforward to show from Eq.~\eqref{F2} that ${\cal E}^{(0)}_{\mu \nu}$, which is defined to be ${\cal E}_{\mu \nu}$ for $K=0$, is symmetric and traceless, and can be expressed as
\begin{eqnarray}\label{F4}
 \kappa~{\cal E}^{(0)}_{\mu \nu}=\frac 14 h_{\mu \alpha,\beta}h_{\nu}{}^{\alpha}{}_{,}{}^{\beta}+\frac 14 \Big(h_{\alpha  \beta,\mu}h^{\alpha \beta}{}_{,\nu}-\frac 12 \eta_{\mu \nu}~h_{ \alpha \beta,\gamma}h^{\alpha \beta}{}_{,}{}^{\gamma}\Big) \nonumber  \\
-\frac 14 \Big(h_{\mu \alpha,\beta}h^{\alpha \beta}{}_{,\nu}+h_{\nu \alpha,\beta}h^{\alpha \beta}{}_{,\mu}-\frac 12 \eta_{\mu \nu}~h_{ \alpha \beta,\gamma}h^{\gamma \alpha}{}_{,}{}^{\beta}\Big)\,.
\end{eqnarray}      
Each of the expressions in Eqs.~\eqref{F3} and~\eqref{F4} consists of the same three parts, but they differ in the overall numerical factors in front of the first and last parts: These are both $\frac 12$ in Eq.~\eqref{F3}, but $\frac 14$ in Eq.~\eqref{F4}. 

For the calculation of the energy flux, we impose the additional gauge condition that $h_{0\mu}=0$. It then follows from Eqs.~\eqref{F3} and\eqref{F4} that
\begin{equation}\label{F5}
{\cal E}^{(0)}_{0k}-t_{0k}=\frac{1}{4\kappa}h_{ki,j}h^{ij}{}_{,0}\,,
\end{equation}
which vanishes in the TT gauge due to the transverse nature of the radiation. For instance, if the spatial axes are so oriented that plane waves propagate to the receiver along the $x^3$ axis, then $k=3$ in Eq.~\eqref{F5} and $h_{3i}=0$ in the TT gauge (see Appendix B). We therefore conclude that in the absence of nonlocality, the flux of gravitational radiation energy will be the same as in standard GR, in agreement with previous results~\cite{SSW}.

Let us next consider the contribution of the nonlocal terms in Eq.~\eqref{F2}. In the Fourier domain, the ratio of the nonlocal term to the corresponding local term is given by $\hat{K} =- \hat{R}/(1+\hat{R})$ 
in accordance with Eq.~\eqref{36}. However, as explained in section V, $|\hat{R}|\sim (80\pi)^{-1}(\lambda/\lambda_0)^2$, which in view of Eq.~\eqref{60} is completely negligible in comparison to unity for radiation of dominant frequency $\gtrsim 10^{-8}$ Hz. We therefore conclude that the nonlocal contribution to the energy-momentum tensor of gravitational radiation can be ignored for gravitational waves that may be detectable in the foreseeable future. 

\section{Discussion} 

Experimental efforts are under way to detect gravitational waves in the frequency range $\gtrsim 10^{-8}$ Hz. We have shown that the treatment of such linearized gravitational waves---namely, their generation, propagation and detection---in nonlocal general relativity essentially reduces to that of general relativity. This circumstance is due to the fact that gravitational waves in such a frequency range have wavelengths that are much smaller than the characteristic scale associated with nonlocality, namely, $\lambda_0=10$ kpc. For the treatment of gravitational waves of frequency $\lesssim c/\lambda_0 \approx 10^{-12}$ Hz, however, nonlocality is expected to play a significant role. 

\begin{acknowledgments}
B.~M. is grateful to F.~W.~Hehl and Yu.~N.~Obukhov for valuable discussions. 
\end{acknowledgments}

\appendix{}
\section{Causal Kernels}\label{appA}

The purpose of this appendix is to present the chain of arguments that lead us to the conclusion that  the causal convolution kernel of linearized nonlocal gravity has a reciprocal causal convolution kernel.

\subsection{causality}

The convolution kernel of linearized nonlocal gravity, $K(x-y)$, is causal; that is, it is zero unless $x-y$ is a future directed timelike or null vector in Minkowski spacetime. This means that  $x^0>y^0$ and 
\begin{equation}\label{A1}
\eta_{\alpha \beta}(x^\alpha-y^\alpha)(x^\beta-y^\beta) \le0\,. 
\end{equation}
Thus
\begin{equation}\label{A2}
(t_x-t_y)^2-\frac{1}{c^2}|\mathbf{x}-\mathbf{y}|^2\ge0 
\end{equation}
and with $t_x>t_y$, the causality requirement reduces to
\begin{equation}\label{A3}
x^0-y^0\ge |\mathbf{x}-\mathbf{y}|\,
\end{equation}
or that 
\begin{equation}\label{A4}
K(x-y)\propto H(x^0-y^0-|\mathbf{x}-\mathbf{y}|)\,. 
\end{equation}
The question is whether the reciprocal kernel satisfies a similar relation as Eq.~\eqref{A4}. To show this using the Fourier transform method is not simple due to the existence of retardation effects---see section IV; however, as discussed below, one can employ the Liouville-Neumann method of successive substitutions~\cite{Tr} to demonstrate that under physically reasonable conditions the reciprocal kernel is causal as well~\cite{MR, FV}, so that reciprocity holds for causal convolution kernels.

\subsection{Volterra algebra}

The convolution property is independent of causality. To emphasize this point, we start in this subsection with a general causal kernel $K(x,y)$. A causal kernel function that is continuous on causally ordered sets in Minkowski spacetime is called a \emph{Volterra kernel}. Volterra kernels have some interesting properties that we briefly mention here. 

We define the product of Volterra kernels $K$ and $K'$ to be 
\begin{equation}\label{A5}
V(x,y) =  \int K(x,z) K'(z,y)d^4z\,. 
\end{equation}
The integrand here     is nonzero only if
\begin{equation}\label{A6}
x^0-z^0\ge |\mathbf{x}-\mathbf{z}|\,,\qquad  z^0-y^0\ge |\mathbf{z}-\mathbf{y}|\,.
\end{equation}
Summing these two conditions leads to 
\begin{equation}\label{A7}
x^0-y^0\ge |\mathbf{x}-\mathbf{z}|+ |\mathbf{z}-\mathbf{y}|\ge |\mathbf{x}-\mathbf{y}|
\end{equation}
by the triangle inequality. Thus $V$ is a Volterra kernel as well. The space of Volterra kernels forms an algebra over the causally ordered events in Minkowski spacetime. It is important to note that the integration in Eq.~\eqref{A5} takes place over a domain ${\cal D} (x,y)$, which is the  region in spacetime that is \emph{bounded} by the past light cone of event $x$ and the future light cone of event $y$, as depicted schematically in Figure 1. For a Volterra kernel $K(x,y)$, it proves useful to define iterated (Volterra) kernels $K_n(x,y)$ for $n=1,2,3,...,$ such that $K_1(x,y)=-K(x,y)$ and 
\begin{equation}\label{A8}
K_{n+1}(x,y) =  \int_{{\cal D}(x,y)} K_n(x,z) K_1(z,y)d^4z\,. 
\end{equation}

\begin{figure}
\centerline{\includegraphics[width=3in]{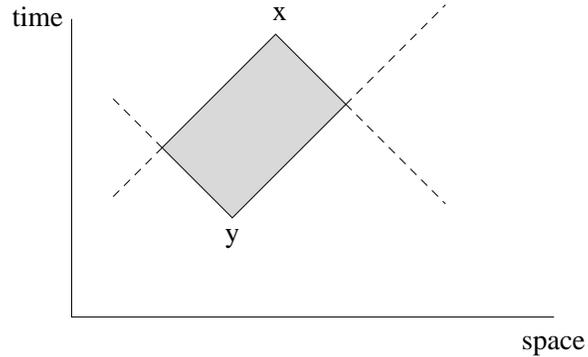}}
\caption{Schematic plot indicating the finite shaded domain $\mathcal{D}(x,y)$ in spacetime. It is the region common to the light cone that has its vertex at event $x$ and the light cone that has its vertex at event $y$.}\label{fig1}
\end{figure}

It is straightforward to check by a simple change of variable in the integral in Eq.~\eqref{A5} that if $K(x,y)$ and $K'(x,y)$ are
convolution kernels and thus just a function of $x-y$, then $V$ is a convolution kernel as well.  Moreover, the order of the terms in the integrand of Eq.~\eqref{A5} is immaterial in this case, so that $V$ is also the product of $K'$ and $K$. Indeed, \emph{convolution Volterra kernels} form a \emph{commutative subalgebra} of the Volterra algebra. 

Next, consider Volterra kernels $K$ and ${\cal A}$; we wish to find a Volterra kernel ${\cal B}$ such that
\begin{equation}\label{A9}
{\cal B}(x,y)+ \int_{{\cal D}(x,y)} K(x,z) {\cal B}(z,y)d^4z = {\cal A}(x,y)\,.
\end{equation}
This generalized Volterra equation of the second kind has a \emph{unique solution} in accordance with the theorem of M. Riesz~\cite{MR, FV} given by
\begin{equation}\label{A10}
{\cal A}(x,y)+ \int_{{\cal D}(x,y)} R(x,z) {\cal A}(z,y)d^4z = {\cal B}(x,y)\,,
\end{equation}
where
\begin{equation}\label{A11}
R(x,y) = \sum_{n=1}^{\infty} K_n(x,y)\,.
\end{equation}
The Neumann series here converges uniformly on bounded domains and the reciprocal kernel $R$ is a Volterra kernel. The proof employs generalized Riemann-Liouville kernels and has been sketched in Ref.~\cite{FV}. The original study of M. Riesz~\cite{MR} was set in a more general context; in our brief presentation here, we have followed the work of Faraut and Viano~\cite{FV}.

\subsection{$L^1$ and $L^2$ Volterra convolution kernels}

Our aim now is to reduce the generalized Volterra integral Eqs.~\eqref{A9} and~\eqref{A10} to Volterra integral Eqs.~\eqref{32} and~\eqref{33}. To this end, we restrict our attention to Volterra \emph{convolution} kernels that are $L^1$ and $L^2$ functions on spacetime. Indeed, these properties need to hold only within and on the future light cone as a consequence of the causality requirement.  

Let $f(x)$ be any continuous $L^1$ function over spacetime and define 
\begin{equation}\label{A12}
{\cal F}(x)=\int {\cal A}(x-y) f(y)d^4y\,, \qquad  {\cal G}(x)= \int {\cal B}(x-y) f(y)d^4y\,.
\end{equation}
Then, multiplying Eqs.~\eqref{A9} and~\eqref{A10} for $L^1$ and $L^2$ \emph{convolution} kernels by $f(y)$, integrating over spacetime and employing Young's inequality for convolutions, we recover Eqs.~\eqref{32} and~\eqref{33} such that ${\cal F}(x)$ and ${\cal G}(x)$ are continuous $L^1$ functions over spacetime. Furthermore, it is a consequence of Minkowski's integral inequality that if $f$ is $L^1$ and ${\cal A}$ is $L^2$, then their convolution is $L^2$. Therefore, ${\cal F}(x)$ and ${\cal G}(x)$ are $L^2$ functions over spacetime as well. In particular, for our $L^1$ and $L^2$ \emph{convolution} reciprocal kernel we have that
\begin{equation}\label{A13}
R(x-y) \propto H(x^0-y^0- |\mathbf{x}-\mathbf{y}|)\,.
\end{equation}

\section{Gauge Conditions}\label{appB}

The various gauge conditions imposed in this paper are essentially the same as in standard general relativity and their discussion is therefore relegated to this appendix. 

We begin with Eq.~\eqref{23} and note that in general 
\begin{equation}\label{B1}
\overline{h'}^{\mu \nu}{}_{,\nu}=\overline{h}^{\mu \nu}{}_{,\nu}+\square \, \epsilon^\mu\,.\end{equation}
Let us first find the gauge functions $\epsilon^\mu$ for which the transverse gauge condition $\overline{h'}^{\mu \nu}{}_{,\nu}=0$ is satisfied. It follows from Eq.~\eqref{B1} that
\begin{equation}\label{B2}
\square \, \epsilon^\mu=-\overline{h}^{\mu \nu}{}_{,\nu}\,.
\end{equation}
Appropriate solutions of this standard inhomogeneous wave equation can be found to ensure that 
 the transverse gauge condition is indeed satisfied. 
 
Next, we start from such trace-reversed potentials $\overline{h}_{\mu \nu}$ that satisfy $\overline{h}^{\mu \nu}{}_{,\nu}=0$, but then note from Eq.~\eqref{B1} that such potentials are not unique. A further gauge transformation leads to potentials $\overline{h'}_{\mu \nu}$ that still satisfy $\overline{h'}^{\mu \nu}{}_{,\nu}=0$, provided the four gauge functions $f^\mu(x)$ satisfy the wave equation 
\begin{equation}\label{B3}
 \square \, f^\mu(x)=0\,.
\end{equation}

Let us further assume that  $\square\, \overline{h}_{0 \mu}=0$ for \emph{plane} gravitational waves as in section V. We wish to show that the remaining four gauge functions $f_\mu$ can now be so chosen as to set $\overline{h'}_{0 \mu}=0$ and $\overline{h'}=0$ as well. To this end, let us choose the spatial inertial coordinate system $(x^1, x^2, x^3)$ such that the direction of propagation of the waves from the source to the detector coincides, for instance, with the positive $x^3$ direction. This simplifies the analysis without any loss in generality. Therefore, near the detector, the wave front can be locally approximated by a plane such that $\overline{h}_{0 \mu}=\overline{h}_{0 \mu}(\zeta)$, where $\zeta:=x^3-x^0$. Setting $\overline{h'}_{0 \mu}=0$ in Eq.~\eqref{23}, $\overline{h'}=0$ in Eq.~\eqref{24} and replacing $\epsilon_\mu$ in these equations by $f_\mu (\zeta)$, we find
\begin{equation}\label{B4}
\frac{df_1}{d\zeta}= \overline{h}_{01}\,, \qquad \frac{df_2}{d\zeta}= \overline{h}_{02}\,,             
\end{equation}
\begin{equation}\label{B5}
\frac{d(f_0-f_3)}{d\zeta}=- \overline{h}_{03}=\overline{h}_{00}\,, \qquad \frac{d(f_0+f_3)}{d\zeta}= \frac{1}{2}\overline{h}\,.            
\end{equation}
Let us note that $\overline{h}_{00}+\overline{h}_{03}=0$ in Eq.~\eqref{B5} is consistent with the transverse gauge condition $\overline{h}^{0 \nu}{}_{,\nu}=0$, which implies that $d(\overline{h}_{00}+\overline{h}_{03})/d\zeta=0$. This relation can be integrated and the integration constant set to zero, as the presence of a nonzero constant here would be inconsistent with the fact that these potentials originate from the distant source of gravitational waves. It is thus evident that $f_\mu$ can be so chosen as to render $h'_{\mu \nu}$ purely spatial and traceless as well. Moreover, we note in connection with the treatment in section VI that in this procedure $\overline{h}_{11}- \overline{h}_{22}$ and $ \overline{h}_{12}$ remain invariant; that is, 
$\overline{h'}_{11}- \overline{h'}_{22}=\overline{h}_{11}- \overline{h}_{22}$ and $ \overline{h'}_{12}= \overline{h}_{12}$.

Finally, let us assume that in addition to the gauge conditions already discussed above, $h_{ij}$ is such that $\square\, h_{ij}=0$. Again, near the receiver in the wave zone, the spherical gravitational waves associated with these potentials locally behave as plane waves and, as before, we can assume that $h_{ij}=h_{ij}(\zeta)$. Then, $h^{ij}{}_{,j}=0$ implies that $dh_{i3}/d\zeta=0$ and hence $h_{i3}(\zeta)=0$. In this way, we recover the TT gauge of GR, where the two independent states of gravitational radiation are given by $h_{11}=-h_{22}$ and $h_{12}=h_{21}$.

\section{Fourier Transform of the Reciprocal Kernel}\label{appC}

The purpose of this appendix is to compute 
\begin{equation}\label{C1}
\hat{R}(\omega, \mathbf{k})= \int R(t, \mathbf{x}) e^{i\omega t-i\mathbf{k} \cdot \mathbf{x}}~dt d^3x\,,
\end{equation}
where in accordance with Eq.~\eqref{41},
\begin{equation}\label{C2}
R(t, \mathbf{x})= H(t-r)Ae^{-At}q(r)\,,
\end{equation}
$A>0$, $r=|\mathbf{x}|$ and $q$ is either $q_1$ or $q_2$ given explicitly in section III. Integrating over the temporal interval $t: r \to \infty$, we find
\begin{equation}\label{C3}
\hat{R}(\omega, \mathbf{k})= \frac{A}{A-i\omega}\int q(r) e^{-i\mathbf{k} \cdot \mathbf{x}}J(\omega, r) d^3x\,,
\end{equation}
where
\begin{equation}\label{C4}
J(\omega, r):=e^{-(A-i\omega)r}\,
\end{equation}
is such that $|J(\omega, r)|\le1$. Neglecting retardation in Eq.~\eqref{C2}, which is the simple approach adopted in section IV, amounts to setting $J(\omega, r)$ equal to unity in Eq.~\eqref{C3}.

Using spherical polar coordinates and the standard result that 
\begin{equation}\label{C5}
\int_0^{\pi}e^{-i |\mathbf{k}| r \cos \theta}\sin \theta d\theta=2 \frac{\sin(|\mathbf{k}| r)}{|\mathbf{k}| r}\,, 
\end{equation}
we find
\begin{equation}\label{C6}
\hat{R}(\omega, \mathbf{k})= \frac{4\pi A}{(A-i\omega)|\mathbf{k}|}\int_0^{\infty} rq(r) \sin(|\mathbf{k}|r)J(\omega, r) dr\,.
\end{equation}

To proceed further, we assume that $\omega=|\mathbf{k}|$ corresponds to frequencies $\gtrsim 10^{-8}$ Hz and $A=\alpha=(10~\lambda_0)^{-1}$, so that $A/\omega \lesssim 10^{-6}$. To calculate the integral in Eq.~\eqref{C6}, let us note that $\hat{R}$ for $\omega=|\mathbf{k}|$ can be written as 
\begin{equation}\label{C7}
\hat{R}= -\frac{2\pi A}{(\omega^2+i\omega A)}\int_0^{\infty} rq(r)e^{-Ar}\Big(1-e^{2i\omega r}\Big) dr\,.
\end{equation}
Let us now use $q_1$ and $q_2$, given respectively by Eqs.~\eqref{30} and~\eqref{31}, for $q$ and  rewrite this expression in terms of dimensionless quantities
\begin{equation}\label{C8}
X:=Ar, \qquad \beta:=\frac{\omega}{A} \gtrsim 10^{6}, \qquad w:=\alpha a\,.
\end{equation}
Then, 
\begin{equation}\label{C9}
\hat{R}= -\frac{5}{\beta^2(1+i/\beta)} [I^{(0)}(w)-I^{(\beta)}(w)]\,,
\end{equation}
where $I^{(0)}$ is formally obtained from $I^{(\beta)}$ for $\beta=0$ and $0<w\ll1$. Moreover,
\begin{equation}\label{C10}
I^{(\beta)}_1=\int_0^{\infty} \frac{X(X+1+w)}{(X+w)^2}~e^{-2(1-i\beta)X}dX, \qquad I^{(\beta)}_2=\int_0^{\infty} \frac{X+1+w}{X+w}~e^{-2(1-i\beta)X}dX\,,
\end{equation}
depending respectively on whether $q_1$ or $q_2$ is used for $q$ in Eq.~\eqref{C7}. Let us recall here that $w=10^{-4}$ for the two examples involving $q_1$ and $q_2$ that were worked out numerically in Ref.~\cite{21}.

It follows from the Riemann-Lebesgue lemma that $I^{(\beta)} \to 0$ as $\beta \to \infty$. Indeed, by using $\exp{(2i\beta X)}=(2i\beta)^{-1}d(\exp{(2i\beta X)})/dX$ in $I^{(\beta)}$, and repeated integrations by parts, it is possible to express $I^{(\beta)}$ as an asymptotic series in powers of $1/\beta$ as $\beta \to \infty$. Furthermore, $I^{(0)}$ can be expressed in terms of the exponential integral function---see, for instance, page 311 of Ref.~\cite{G+R}. We find
\begin{equation}\label{C11}
I^{(0)}_1=-\frac{1}{2}-(1+w)e^{2w} {\rm Ei} (-2w)\,, \qquad I^{(0)}_2=\frac{1}{2}-e^{2w} {\rm Ei} (-2w)\,.
\end{equation}
Thus for $0<w\ll1$, we have 
\begin{equation}\label{C12}
I^{(0)}_1\approx -\frac{1}{2}-[{\cal C}+\ln{(2w)}]\,, \qquad I^{(0)}_2\approx I^{(0)}_1+1\,,
\end{equation}
since for $0<x\ll 1$, ${\rm Ei}(-x)\approx {\cal C}\, +\, \ln x$, where ${\cal C}=0.577...$ is the Euler-Mascheroni constant---see page 927 of Ref.~\cite{G+R}. More precisely, for $w=10^{-4}$, $I^{(0)}_1\approx 7.443$ and $I^{(0)}_2\approx 8.442$, while for $w=10^{-3}$, $I^{(0)}_1\approx 5.156$ and $I^{(0)}_2\approx 6.151$.

We are now in a position to compare the above exact expression for $|\hat{R}|$ in the $\omega=|\mathbf{k}|$ case with the estimate used in section V, namely, $|\hat{R}|\sim 5\pi/\beta^2$. We find that for $\beta$ as in Eq.~\eqref{C8}, our estimate for $|\hat{R}|$ is smaller than the more exact value calculated here by a factor of about 2 or 3 for $w$ around $10^{-3}$ or $10^{-4}$, respectively.


\begin{thebibliography}{99}

\bibitem{1} 
A.~Einstein, \textit{The Meaning of Relativity}  (Princeton University Press, Princeton, NJ, 1955).

\bibitem{2} 
B.~Mashhoon, Phys. Rev. Lett. \textbf{61}, 2639 (1988).

\bibitem{3}  
B.~Mashhoon, Phys. Lett. A \textbf{143}, 176 (1990).

\bibitem{4} 
B.~Mashhoon, Phys. Lett. A \textbf{145}, 147 (1990).

\bibitem{5a} 
 B.~Mashhoon, Lect. Notes Phys. \textbf{514}, 269 (1998); arXiv: gr-qc/0003014.
 
\bibitem{5b}
B.~Mashhoon, in
\textit{Relativity in Rotating Frames}, edited by G.~Rizzi and M.~L.~Ruggiero (Kluwer Academic
Publishers, Dordrecht, 2004), pp. 43--55; arXiv: gr-qc/0303029.


\bibitem{6} 
N.~Bohr and L.~Rosenfeld, K. Dan. Vidensk. Selsk. Mat. Fys. Medd. \textbf{12}, No. 8 (1933);\\ translated in \textit{Quantum Theory and Measurement}, edited by J.~A.~Wheeler and W.~H.~Zurek (Princeton University Press, Princeton, NJ, 1983).
	
\bibitem{7} 
N.~Bohr and L.~Rosenfeld, Phys. Rev. \textbf{78}, 794 (1950).

\bibitem{8} 
B.~Mashhoon, % {\it Nonlocal special
                                % relativity,} 
Ann.\ Phys.\ (Berlin) {\bf 17}, 705 %--727
(2008); arXiv: 0805.2926 [gr-qc].

\bibitem{9}
 F.~W.~Hehl, P.~von der Heyde, G.~D.~Kerlick and J.~M.~Nester, Rev. Mod. Phys.  {\bf 48}, 393 (1976). 
 
 \bibitem{10}
 F.~W.~Hehl, J.~Nitsch and P.~Von der Heyde, ``Gravitation and the Poincar\'e Gauge Field Theory with Quadratic Lagrangian'', in {\it
    General Relativity and Gravitation}, edited by A.~Held
 (Plenum, New York, 1980), Vol. 1, pp.\ 329--355.


 
\bibitem{11} 
J.~Nitsch and F.~W.~Hehl, %{\it Translational gauge
    %theory of gravity: Post-Newtonian approximation and spin
    %precession,} 
 Phys.\ Lett. B  {\bf 90}, 98 %--102 
  (1980).


\bibitem{12}
M.~Blagojevi\'c and F.~W.~Hehl,
{\it Gauge Theories of Gravitation}
(Imperial College Press, London, UK, 2012).

\bibitem{13}
E.~L.~Schucking and E.~J.~Surowitz, arXiv: gr-qc/0703149v2.

\bibitem{14}
R.~Aldrovandi and J.~G.~Pereira, 
\textit{Teleparallel Gravity: An Introduction} (Springer, New York, 2013).

\bibitem{15}  
F.~W.~Hehl and Yu.~N.~Obukhov, {\it Foundations of
    Classical Electrodynamics: Charge, Flux, and Metric} (Birkh\"auser, Boston, MA, 2003).

\bibitem{16} 
U.~Muench, F.~W.~Hehl and B.~Mashhoon, %{\it
 %   Acceleration-Induced Nonlocal Electrodynamics in Minkowski
 %   Spacetime,} 
  Phys.\ Lett.\ A  {\bf 271}, 8 (2000); arXiv: gr-qc/0003093.
  
\bibitem{Sy}
 J.~L.~Synge, {\it Relativity: The General Theory} (North-Holland, Amsterdam, 1971).
    

    
\bibitem{17} F.~W.~Hehl and B.~Mashhoon, Phys.\ Lett.\ B {\bf
    673}, 279 (2009); arXiv: 0812.1059 [gr-qc].

\bibitem{18} F.~W.~Hehl and B.~Mashhoon, Phys.\ Rev.\ D {\bf
    79}, 064028 (2009); arXiv: 0902.0560 [gr-qc].

\bibitem{19} H.-J.~Blome, C.~Chicone, F.~W.~Hehl and B.~Mashhoon, Phys.\ Rev.\ D {\bf
    81}, 065020 (2010); arXiv: 1002.1425 [gr-qc].
    
\bibitem{20} B.~Mashhoon, ``Nonlocal Gravity", in \emph{Cosmology and Gravitation}, edited by M. Novello and S. E. Perez Begliaffa (Cambridge Scientific Publishers, UK, 2011), pp. 1--9; arXiv: 1101.3752 [gr-qc].

\bibitem{21} C.~Chicone and B.~Mashhoon, J.\ Math.\ Phys. {\bf
    53}, 042501 (2012); arXiv: 1111.4702 [gr-qc].
    
\bibitem{AGP}
V.~C.~de Andrade, L.~C.~T.~Guillen and J.~G.~Pereira, 
%Gravitational energy-momentum density in teleparallel gravity, 
Phys.\ Rev.\ Lett. {\bf 84}, 4533 (2000).
    
    
\bibitem{SS} M.~Schweizer and N.~Straumann, Phys.\ Lett.\ A  {\bf 71}, 493 (1979). 
        
\bibitem{SSW} M.~Schweizer, N.~Straumann and A.~Wipf,  Gen.\ Relativ.\ Gravit.  {\bf 12}, 951 (1980). 

\bibitem{MGH} 
U.~Muench, F.~Gronwald and F.~W.~Hehl, Gen.\ Relativ.\ Gravit.  {\bf 30}, 933 (1998). 

\bibitem{Ri} K.~Riles, arXiv: 1209.0667 [hep-ex].
    
\bibitem{Tr} F.~G.~Tricomi, {\it Integral Equations} (Interscience, New York, 1957).


\bibitem{BC1} D.~Clowe \emph{et al.}, Astrophys.\ J.\ Lett.\ {\bf
    648}, L109 %--L113
  (2006).

\bibitem{BC2} D.~Clowe, S.~W.~Randall and M.~Markevitch, Nucl.\ Phys.\ B
  (Proc. Suppl.) {\bf 173}, 28 (2007). 
  
 \bibitem{RF} 
V.~C.~Rubin and W.~K.~Ford, Astrophys.\ J.  {\bf 159}, 379 (1970).  

\bibitem{RW}
 M.~S.~Roberts and R.~N.~Whitehurst, Astrophys.\
  J.  {\bf 201}, 327 (1975).

\bibitem{SR} 
Y.~Sofue and V.~Rubin, %{\it Rotation curves of spiral
%    galaxies,} 
Annu.\ Rev.\ Astron.\ Astrophys. {\bf 39}, 137 %--174
    (2001).
     
\bibitem{T} J.~E.~Tohline, in {\it IAU Symposium 100,
    Internal Kinematics and Dynamics of Galaxies,} edited by
  E.~Athanassoula (Reidel, Dordrecht, 1983), p.~205.
  
\bibitem{K} J.~R.~Kuhn and L.~Kruglyak,   Astrophys.\ J. {\bf 313}, 1 (1987).

\bibitem{B} J.~D.~Bekenstein, in {\it Second Canadian
    Conference on General Relativity and Relativistic Astrophysics}, edited by
  A.~Coley, C.~Dyer and T.~Tupper (World Scientific, Singapore,
  1988), p.~68.

\bibitem{TF}
 R.~B.~Tully and J.~R.~Fisher, Astron.\ and Astrophys. {\bf 54}, 661 (1977).

\bibitem{Me} H.~Meyer \emph{et al.}, Gen.\ Relativ.\ Gravit.  {\bf 44}, 2537 (2012). 

\bibitem{MR}
M.~Riesz, Acta\ Math. {\bf 81}, 1 (1949).

\bibitem{FV}
J.~Faraut and G.~A.~Viano, J.\ Math.\ Phys. {\bf 27}, 840 (1986).

\bibitem{OPR}
Yu.~N.~Obukhov, J.~G.~Pereira and G.~F.~Rubilar, 
%On the energy transported by exact plane gravitational-wave solutions, 
Class.\ Quantum\ Grav. {\bf 26}, 215014 (2009).
        
 \bibitem{Mn} 
B.~Mashhoon, Astrophys.\ J.  {\bf 223}, 285 (1978). 

   
\bibitem{G+R} I.~S.~Gradshteyn and I.~M.~Ryzhik, {\it Table of
    Integrals, Series and Products} (Academic Press, New York, 1980).  
    
 
                
\end{thebibliography}
\end{document}